\documentclass{elsart}

\usepackage{natbib}

\usepackage{amssymb}
\usepackage {graphicx}
\usepackage{placeins}
\usepackage{longtable,array,tabularx}
\begin{document}

\begin{frontmatter}



\title{Merger driven ULIRG-QSO evolution:\\ The case of 3C~48}

\author[cologne]{J. Zuther}
\author[hawaii]{M. Krips}
\author[chile]{J. Scharw\"achter}
\author[cologne]{A. Eckart}
\address[cologne]{I. Physikalisches Institut, Universit\"at zu K\"oln, Z\"ulpicher Str. 77, 50937 K\"oln, Germany}
\address[hawaii]{ Smithsonian Astrophysical Observatory (SAO), Submillimeter Array (SMA), 645, North A'Ohoku Place, Hilo, HI, 96720, USA}
\address[chile]{European Southern Observatory, Alonso de Cordova 3107 Vitacura
Casilla 19001, Santiago 19, Chile }

\begin{abstract}
The QSO 3C~48 and its host galaxy constitute a nearby template object of the proposed merger-driven evolutionary sequence from ULIRGs to QSOs. In this contribution multi-wavelength observations and N-body simulations studying the structural and compositional properties of this late-stage major merger will be presented. Key questions addressed will be the nature of the apparent second nucleus 3C~48A and absence of a counter tidal tail. The results will be used to review the role of 3C~48 in the ULIRG-QSO evolutionary scenario.
\end{abstract}


\end{frontmatter}

\section{Overview}
\label{overview}
Revealing the physical properties of quasi-stellar objects (QSOs) and their host galaxies is one key to understand how the overall galaxy structure affects the extreme activity in active galactic nuclei (AGN) and how in turn the QSO and its black hole influence the surrounding galaxy structure. 
Galaxy interaction is regarded to be one of the processes that can lead to the (re-)activation of QSOs.
A prime example of this phenomenon is the radio source 3C~48, one of the first quasars to be optically identified \citep{matthews1963}. Its unusually large and bright host galaxy comprises a young stellar population \citep{kristian1973, boroson1982,boroson1984}. Host galaxy properties like the existence of a second bright compact component, 3C~48A, about $1''$ northeast of the QSO \citep{stockton1991,zuther2004}, the tail-like extension to the northeast \citep{canalizo2000}, and the richness of 3C~48 in molecular gas \citep{scoville1993,wink1997} have been used as arguments that the activity in 3C~48 is triggered by a recent major merger. In this contribution we will briefly address the following questions, that are important in order to get a more comprehensive understanding of the relevant physical feedback in this object:
\begin{itemize}
\item Is 3C~48A the nucleus of the companion?
\item Where is the counter tidal tail, commonly observed in major mergers?
\item How does 3C~48 fit into the merger driven ULIRG-QSO evolutionary sequence?
\end{itemize}

\section{The second nucleus: molecular and continuum emission}
\begin{figure}[h!]
\begin{center}
\includegraphics[width=13cm]{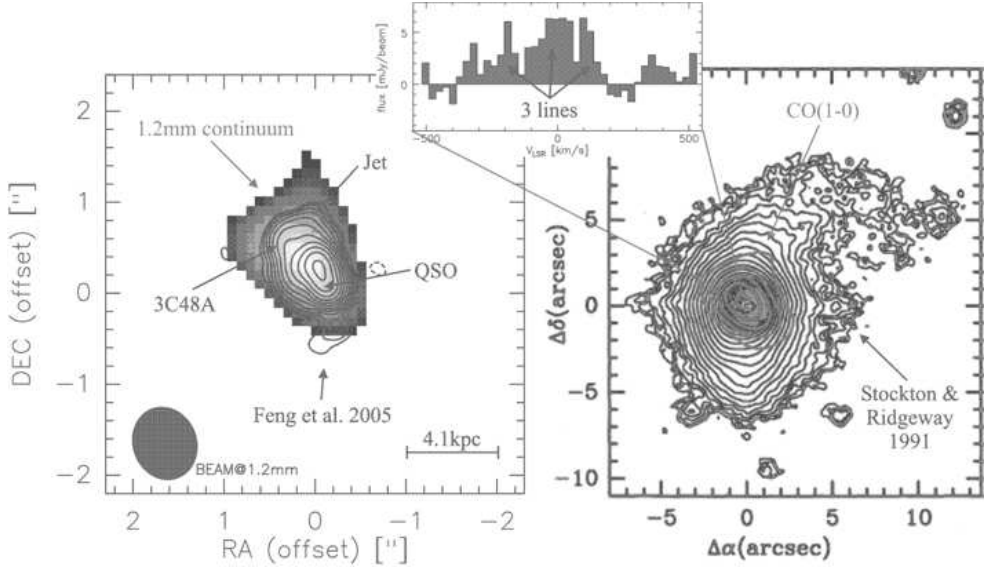}
\caption{
\emph{Left panel:} Millimeter continuum image overlaid with a MERLIN 1.6 GHz map \citep[black contours;][]{feng2005}. The morphology of the mm continuum can be fitted with three Gaussian components: (i) the QSO nucleus, (ii) the radio jet, and (iii) 3C~48A. \emph{Right panel:} optical image \citep[black contours;][]{stockton1991} overlaid with our CO(1-0) measurement (thicker, lighter contours). The spectrum reveals that the CO(1-0) line decomposes into two line systems. The central line, which also contains the main part of the flux, corresponds to 3C~48A. The second gas reservoir (red- and blue-shifted lines) is concentrated around the QSO nucleus. The velocity gradients seen in both systems are almost perpendicular to each other, indicating the presence of two rotating disks of molecular gas. 
}
\label{fig:fig1}
\end{center}
\end{figure}

Our near-infrared (NIR) imaging and spectroscopy result in a first detection of 3C~48A in $J$, $H$, and $Ks$ \citep{zuther2004}. Its reddening in the $H-Ks$ direction indicates the presence of warm dust, either heated by star formation, the active nucleus, or both.
Recent mm-continuum and CO(1-0) observations with the Plateau de Bure interferometer (PdBI), in combination with existing PdBI data \citep{wink1997}, back up the idea of the existence of a second nucleus 3C~48A \citep[][see Fig. \ref{fig:fig1}]{krips2005}. We find indications for two distinct gas reservoirs. One is located at the position of the QSO. But most of the molecular emission accounts for the second reservoir located around 3C~48A. The total molecular gas mass is of the order of $10^{10}$ M$_\odot$, which is typical for advanced merger ULIRGs, because they generally contain the gas of the two gas-rich precursor galaxies.

\section{The missing counter tidal tail and the gas distribution}
\label{sec:countertail}
\begin{figure}[h!]
\begin{center}
\includegraphics[width=8cm]{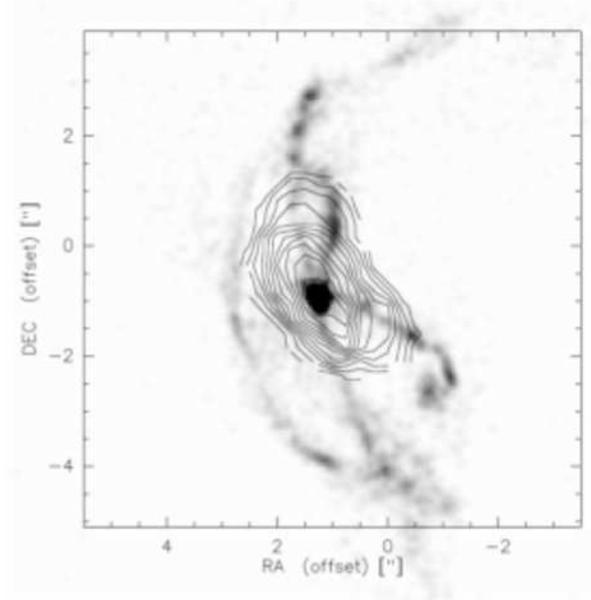}
\caption{
Overlay of the integrated CO(1-0) intensity measured for 3C~48 \citep[contours;][]{krips2005} on the gas density distribution by 3C~48 model \citep[grey-scale image;][]{scharw2005}.
}
\label{fig:fig2}
\end{center}
\end{figure}

Using multi-particle simulations of the 3C~48 merger \citep{scharw2004} we can offer an explanation to the apparently missing counter tidal tail. This might be due to a projection effect of the complex morphology forming during the merger of tilted disk galaxies. In the model the second tail is located in front of the host galaxy \citep[see Fig. 2 in][]{scharw2004}. It is therefore blended with the bright underlying host and nucleus.
The distribution of the gas component resulting from a smoothed particle hydrodynamics simulation (SPH) shows close resemblance with the observed CO(1-0) distribution (Fig. \ref{fig:fig2}). It displays a high central gas concentration which contains about 50\% of all used gas particles. Gas lanes are also aligned with the north-southwestern extension of the molecular emission. The mere presence of gas inflows in the model suggests a considerable star formation activity which is indeed found in the 3C~48 host by \citet{canalizo2000}.

\section{3C~48 in context}
\label{sec:context}
In the ULIRG-QSO evolutionary scenario proposed by \citet{sanders1988} the gravitational interaction of galaxies can trigger the formation of ULIRGs, which can evolve in an advanced merger stage into a QSO, as they turn on the nuclear activity of the super-massive black holes at their centers.
3C~48 queues in the growing list of powerful AGN in which molecular gas is detected in the two components imagined to constitute the merger. 
The morphology of the merger remnant is complex and depends strongly on the viewing angle. In this context, we can show that the missing counter tidal tail possibly lies in front of the galaxy.
These new findings support the rank of 3C~48 in the above evolutionary scheme as a QSO at a late major-merger stage, but still emitting most of its luminosity in the FIR which on the other hand is characteristic for ULIRGs.





\begin{thebibliography}{}
\bibitem[Boroson \& Oke(1982)]{boroson1982}
Boroson, T. A., Oke, J. B. 1982, Nature, 296, 397

\bibitem[Boroson \& Oke(1984)]{boroson1984}
Boroson, T. A., Oke, J. B. 1984, ApJ, 281, 535

\bibitem[Canalizo \& Stockton(2000)]{canalizo2000}
Canalizo, G., Stockton, A. 2000, ApJ, 528, 201

\bibitem[Feng et al.(2005)]{feng2005}
Feng, W. X., et al. 2005, A\&A, 434, 101

\bibitem[Krips et al.(2005)]{krips2005}
Krips, M., Eckart, A., Neri, R., Zuther, J., Downes, D., Scharw\"achter, J. 2005, A\&A, 439, 75

\bibitem[Kristian(1973)]{kristian1973}
Kristian, J. 1973, ApJ, 179, 61

\bibitem[Matthews et al.(1963)]{matthews1963}
Matthews, T. A., Sandage, A. R. 1963, ApJ, 138, 30

\bibitem[Sanders et al.(1988)]{sanders1988}
Sanders, D. B., et al. 1988, ApJ, 325, 74

\bibitem[Scharw\"achter et al.(2004)]{scharw2004}
Scharw\"achter, J., Eckart, A., Pfalzner, S., Zuther, J., Krips, M., Straubmeier, C.  2004, A\&A, 414, 497

\bibitem[Scharw\"achter(2005)]{scharw2005}
Scharw\"achter, J. 2005, PhD thesis, University of Cologne

\bibitem[Scoville et al.(1993)]{scoville1993}
Scoville, N. Z., Padin, S., Sanders, D. B., Soifer, B. T., Yun, M. S. 1993, ApJ, 415L, 75

\bibitem[Stockton \& Ridgway(1991)]{stockton1991}
Stockton, A., Ridgway, S. E. 1991, AJ, 102, 488

\bibitem[Wink, Guilloteau, \& Wilson(1997)]{wink1997}
Wink, J. E., Guilloteau, S., Wilson, T. L. 1997, A\&A, 322, 427

\bibitem[Zuther et al.(2004)]{zuther2004}
Zuther, J., Eckart, A., Scharw\"achter, J., Krips, M., Straubmeier, C. 2004, A\&A, 414, 919

\end{thebibliography}
\end{document}